\documentclass[printer]{aa}
\usepackage{graphicx}
\graphicspath{ {plots/} }
\usepackage[]{natbib}
\usepackage{amsmath}
\usepackage{amssymb}
\usepackage{hyperref}
\usepackage{soul}
\usepackage[dvipsnames]{xcolor}
\usepackage{tikz}
\hypersetup{
    colorlinks=true,
    linkcolor=blue,
    filecolor=magenta,      
    urlcolor=cyan,
    citecolor=blue
}

\bibpunct{(}{)}{;}{a}{}{,} 

\titlerunning{}

\def\teff{\textit{T}_{\text{eff}}}

\def\logg{\text{log}(\textit{g})}

\def\feh{[\text{Fe}/\text{H}]}

\begin{document}

\renewcommand{\arraystretch}{1.5}

\title{Observational constraints on the origin of the elements.\\VIII. Constraining the Barium, Strontium and Yttrium chemical evolution in metal-poor stars}

\author{G.~Guiglion \inst{1, 2, 3},
M.~Bergemann \inst{2},
N.~Storm \inst{2},
J. Lian \inst{4, 2},
G.~Cescutti \inst{5, 6, 7},
A. Serenelli  \inst{8, 9}}

\institute{Zentrum f\"ur Astronomie der Universit\"at Heidelberg, Landessternwarte, K\"onigstuhl 12, 69117 Heidelberg, Germany \\ \email{guiglion@mpia.de}
\and
Max Planck Institute for Astronomy, K\"onigstuhl 17, 69117, Heidelberg, Germany
\and
Leibniz-Institut f\"ur Astrophysik Potsdam (AIP), An der Sternwarte 
16, 14482 Potsdam, Germany
\and
South-Western Institute for Astronomy Research, Yunnan University, Kunming, Yunnan 650091, People’s Republic of China
\and
Dipartimento di Fisica, Sezione di Astronomia, Università di Trieste, Via G. B. Tiepolo 11, 34143 Trieste, Italy
\and 
INAF – Osservatorio Astronomico di Trieste, Via Tiepolo 11, 34143 Trieste, Italy
\and
INFN – Sezione di Trieste, Via A. Valerio 2, 34127 Trieste, Italy
\and
Institute of Space Sciences (ICE, CSIC), 08193, Cerdanyola del Valles, Spain
\and
Institut d'Estudis Espacials de Catalunya (IEEC), 08034, Barcelona, Spain
}
    
\date{Received XXX; accepted XXX}
 
  \abstract
   {The chemical evolution history of slow neutron-capture elements in the Milky Way is still a matter of debate, especially in the metal-poor regime ([Fe/H]<-1).}
   {Recently \citet{Lian2023}, thanks to \emph{Gaia}-ESO spectroscopic data, studied the chemical evolution of neutron-capture elements in the regime [Fe/H]>-1. We aim here to complement this study down to [Fe/H]=-3, and focus on Ba, Y, Sr, and abundance ratios of [Ba/Y] and [Sr/Y], which give comprehensive views on s-process nucleosynthesis channels.}
   {We measured LTE and NLTE abundances of Ba, Y, and Sr in 323 Galactic metal-poor stars using high-resolution optical spectra with high signal-to-noise. We used the spectral fitting code TSFitPy, together with 1D model atmospheres using previously determined LTE and NLTE atmospheric parameters.}
   {We find that the NLTE effects are on the order of $\sim -0.1$ to $\sim 0.2$ dex depending on the element. We find that stars enhanced(deficient) in [Ba/Fe] and [Y/Fe] are also enhanced(deficient) in [Sr/Fe], suggesting a common evolution channel for these three elements. We find that the ratio between heavy and light s-process elements [Ba/Y] varies weakly with [Fe/H] even in the metal-poor regime, consistent with the behaviour in the metal-rich regime. The [Ba/Y] scatter at a given metallicity is larger than the abundance measurement uncertainties. Homogeneous chemical evolution models with different yields prescriptions are unable to accurately reproduce the [Ba/Y] scatter in the low-[Fe/H] regime. Adopting the stochastic chemical evolution model by \citet{Cescutti2014} allows to reproduce the observed scatter in the abundance pattern of [Ba/Y] and [Ba/Sr]. With our observations, we rule out the need for an arbitrary scaling of the r-process contribution as previously suggested by the model authors.}
   {We have showed how important it is to properly include NLTE effects when measuring chemical abundances, especially in the metal-poor regime. This work shows that the choice of the Galactic chemical evolution model (stochastic vs. 1-zone) is key when comparing models to observations. The upcoming large scale spectroscopic surveys such as 4MOST and WEAVE will deliver high quality data of many thousands of metal-poor stars, and this work gives a typical case study of what could be achieved with such surveys.}

\titlerunning{Chemical evolution of Barium, Strontium and Yttrium}
\authorrunning{G. Guiglion, et al.}
\maketitle
 
\section{Introduction}

Neutron-capture elements have been extensively studied in the astronomy community for more than three decades, but the chemical evolution of such type of elements in the Milky Way is still a matter of debate (e.g. \citealt{sneden_2008, Cowan2021}).
Neutron-capture elements are, for instance, essential for tracing the accretion history of the Milky Way (e.g. \citealt{Helmi2020, Matsuno2021}), and its satellite Galaxies (e.g. \citealt{Venn2012}).

\begin{figure*}
\centering
\includegraphics[width=\textwidth]{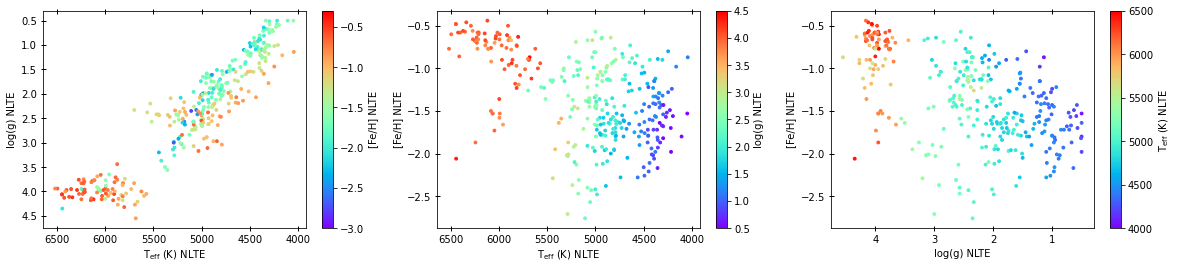}
\caption{Left: Kiel diagram of the 323 stars of the sample
(NLTE $\teff$ vs. $\logg$), colour-coded with NLTE $\feh$.
Middle: NLTE $\feh$ vs. $\teff$ colour-coded with NLTE $\logg$.
Right: NLTE $\feh$ vs. $\logg$ colour-coded with NLTE $\teff$.} 
\label{Kiel}
\end{figure*} 

In particular, the slow-neutron capture (s-process) dominated elements are mainly organised in two peaks \citep{burbidge_1957}. The first peak is located around the magic number 50, and is responsible for the synthesis of the light-s elements Sr, Y, and Zr. The second peak produces Ba, La, Ce, Pr, and Nd around the magic number 80. There exists a third peak as well, that produces Pb. It is common to divide the s-process into a “main” process, a “weak” process and a “strong” process. The \textit{main} s-process occurs in asymptotic giant branch (AGB) stars \citep{Busso1999, bisterzo_2011} while the \textit{weak} s-process is known to occur in massive stars and it produces  nuclei with magic number below 88 \citep{Pignatari2010}. Finally, the \textit{strong} s-process is responsible for about 50\% of solar $^{208}$Pb production by low-metallicity AGB stars \citep{Kaeppeler1982}.

\begin{figure}
\centering
\includegraphics[width=\columnwidth]{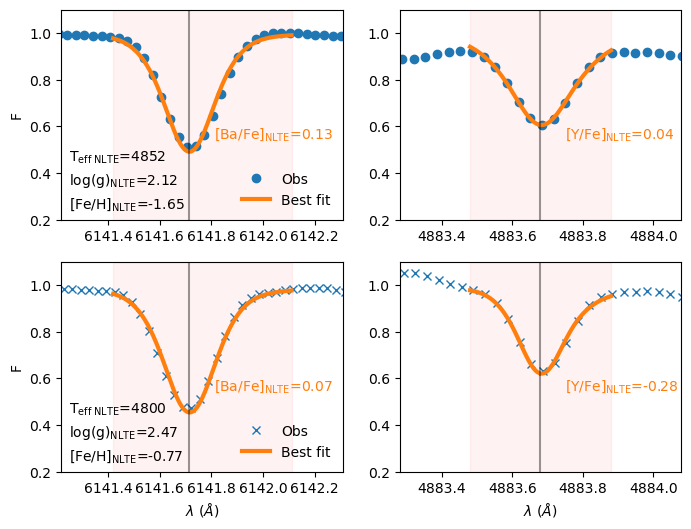}
\caption{Examples of Ba II (left) and Y II (right) lines in the spectra of two red giants with [Fe/H]$_{\mathrm{NLTE}}=-1.65$ (top) and [Fe/H]$_{\mathrm{NLTE}}=-0.77$ (bottom). The red shaded area represents the spectral range over which the line is fitted. The orange curve corresponds to the best line-fit. The black vertical corresponds to the central wavelength of the line.}
\label{spectra}
\end{figure} 

The chemical evolution of neutron-capture elements has been rather well constrained in the Milky Way disc (e.g. \citealt{battistini_bensby_2016, Mishenina2019}). However, not many studies focused on the halo or halo/disc interface. Even with the advent of large-scale spectroscopic surveys (e.g. GALAH; \citealt{buder2019}), neutron-capture abundance measurements are available for only a few tens of very metal-poor stars (e.g. \citealt{Mashonkina2007, Matsuno2021}), as it requires high-quality and high signal-to-noise ratio spectra in the metal-poor regime ($\feh<-1$). Past GCE works also relied on compilations of abundances from various studies, but these may suffer from systematic biases, owing to the fundamental assumption of local thermodynamical equilibrium (LTE)  (e.g. \citealt{Cescutti2014}). High-resolution spectroscopy is indeed a unique technique for determining precise estimates of neutron-capture elemental abundances (e.g. \citealt{delgado_mena_2017, guiglion_2018, Roederer2022}).

Recently \citet{Lian2023} studied the Galactic chemical evolution of Ba and Y using the data from the \emph{Gaia}-ESO large spectroscopic survey. Most stars in the sample of \citet{Lian2023} cover the metallicity range  $-1<\feh<0.5\,$dex, for which \emph{Gaia}-ESO measured and released high-quality chemical abundances of neutron-capture elements. In this letter, we aim to constrain the chemical evolution of Ba, Y, and Sr in the metal-poor
regime ($-3<\feh<-0.5\,$dex), in order to further complement the study of \citet{Lian2023}. We also make one step forward towards a higher accuracy by computing our abundances in the framework of non-local thermodynamic equilibrium (NLTE).

In Section~\ref{sec:data_methodology}, we present the spectroscopic data and spectral analysis. In Section~\ref{sec:results}, we present LTE and NLTE chemical abundance trends of Sr, Y, and Ba, while we confront our observations to Galactic chemical evolution models in Section~\ref{sec:models}. Finally, we present our conclusions in Section~\ref{sec:conclusions}.

\begin{figure*}
\centering
\includegraphics[width=\textwidth]{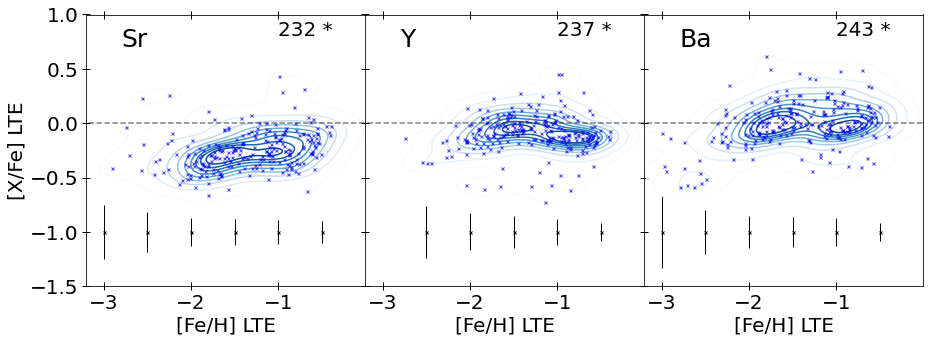}
\includegraphics[width=\textwidth]{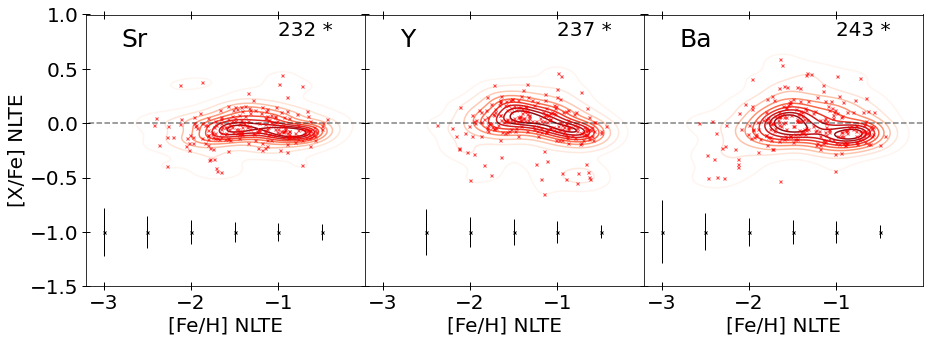}
\caption{Top row: chemical abundances of [Sr/Fe] (left), [Y/Fe] (center), and [Ba/Fe] (right) in LTE as a function of LTE [Fe/H]. We both show individual stars and contour-plot. Black error bars correspond to mean uncertainties $\langle\sigma\rangle$. Bottom row: same as top row, but in NLTE.}
\label{Abund_Ba_Sr_Y}
\end{figure*} 

\section{Data and methodology} \label{sec:data_methodology}

We took advantage of high-resolution spectra of Galactic disc and halo stars from \citet{Ruchti2011}. These targets were originally observed at intermediate resolution by the RAVE survey \citep{steinmetz2006, matijevic2017}. The sample consists of 323 metal-poor stars covering the ranges $4050<\teff<6500\,$K, $0.5<\logg<4.5$, and $-2.8<\feh<-0.4$ (in NLTE; see \figurename~\ref{Kiel}). The data have already been used in our previous studies, e.g, in the analysis of NLTE stellar parameters and metallicities \citep{Ruchti2013}, ages \citep{Serenelli2013}, and NLTE Mg abundances \citep{Bergemann2017a, Bergemann2017b}. We adopted the 1D-LTE and 1D-NLTE atmospheric parameters from \citet{Ruchti2013}.

In order to derive chemical abundances of \ion{Ba}{II}, \ion{Y}{II}, and \ion{Sr}{I}, we used the spectral fitting code
TSFitPy\footnote{https://github.com/TSFitPy-developers/TSFitPy}, which is based on the LTE version of TurboSpectrum \citep{plez_2012} as well as its NLTE extension\footnote{https://github.com/bertrandplez/Turbospectrum\_NLTE} \citep{Gerber2023}. TSFitPy allows to simultaneously fit the abundance, the micro- and macro-turbulence, as well as to apply radial velocity corrections to the data, which are typically needed owing to the lack or realistic convection and turbulent flows in 1D hydrostatic models \citep[see e.g.][]{Dravins2008,Nordlund2009,Meunier2017}.

The model atom of Sr used in this work is based on the model described in \citet{bergemann2012c}, however, it has been updated with new quantum-mechanical data for inelastic  transitions in Sr$+$H collisions \citep{Gerber2023}. The Ba model is that from \citet{Gallagher2020}, and Y model from \citet{Storm2023}.
The atomic transition probabilities come from \citet{Davidson1992} for \ion{Ba}{II} lines, \citet{GarcioCampos1988} for \ion{Sr}{I} lines and \citet{Biemont2011} for Y (see \citep{heiter2021} for more details). \ion{Ba}{II} lines suffer significantly from hyperfine splitting (HFS) and isotopic shifts, and these effects have been included in the calculations as described in \citep{Gallagher2020}. HFS is negligible for \ion{Y}{II} (<0.5 m\AA) and is not included in the linelist. For \ion{Ba}{II}, we adopted the lines at $5853.67$, $6141.71$, $6496.90\,$\AA~that show rather strong features even down to $\feh=-3$. For Sr, we used the strong and unblended spectral line of \ion{Sr}{I} at $4607.33\,$\AA, while two \ion{Y}{II} lines were adopted for the Yttrium measurements ($4883.68$ and $5087.42\,$\AA). It has already been demonstrated in \citet{bergemann2012c} that NLTE provides a robust ionization balance for Sr I and Sr II-based abundances for dwarfs and red giants over the metallicity range relevant to the present work. For Y, the standard validation tests of the NLTE model atom, including validation on metal-poor red giant atmospheres, were presented in \citet{Storm2023}. For the atomic and molecular blends, we took advantage of the comprehensive \emph{Gaia}-ESO survey linelist \citep{heiter2021}. We adopted the extensively used 1D MARCS model atmospheres from \citet{gustafsson_2008}. The solar abundances are taken from \citet{Magg2022}.

We carefully checked by eye the quality of the fitted spectral lines to ensure the robustness of the abundance measurements. Examples of Ba II and Y II best-fit profiles are showed in \figurename~\ref{spectra} for two red giants with different metallicities. For a given element, the error budget $\sigma$ was computed by quadratically summing the line-to-line scatter ($\sigma_{sc}$) to the propagated errors from the three atmospheric parameters ($\teff$, $\logg$, $\feh$) ($\sigma_{atm}$). 

\section{Chemical abundance trends of [Sr/Fe], [Y/Fe], and [Ba/Fe]} \label{sec:results}

In \figurename~\ref{Abund_Ba_Sr_Y}, we show LTE (blue) and NLTE (red) abundances of Sr, Y, and Ba as a function of LTE (top) and NLTE (bottom) metallicity [Fe/H], respectively.

\begin{figure}
\centering
\includegraphics[width=\columnwidth]{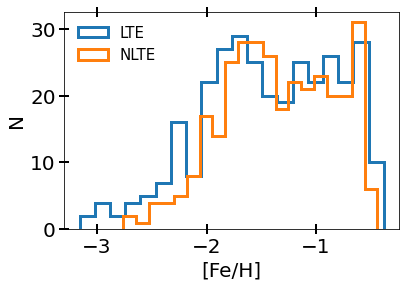}
\caption{LTE and NLTE metallicity distribution of our stellar sample probing the metal-weak Galactic disc and the halo \citep[see][]{Ruchti2013}. The distribution is not uniform and not statistically complete, because of the observational selection function \citep{Ruchti2011}.}
\label{metallicity_distribution_function}
\end{figure}

LTE [Sr/Fe] shows a slightly increasing behaviour with [Fe/H], and the abundance ratios tend to be strongly sub-solar over the entire metallicity range. NLTE [Sr/Fe] is only slightly sub-solar for [Fe/H]>-1, and solar for [Fe/H]<-1. We note a major difference between both LTE and NLTE of about $0.2$ dex in [Sr/Fe]. This is expected, because NLTE effects on the formation of the \ion{Sr}{I} line at 4\,706\,\AA~are positive, as in NLTE the line opacity decreases, leading to weaker lines compared to LTE \citep{bergemann2012c, Hansen2013}. Overall, the stars-by-star scatter is around 0.17 dex for LTE [Sr/Fe], while the scatter drops to 0.13 for NLTE [Sr/Fe], as a result of NLTE effects shrinking the distribution. This is also the case for the individual abundance uncertainties, as illustrated by the black error bars in the bottom of each panel (computed as mean uncertainties in a 0.5\,dex [Fe/H range]). We notice that the overall distribution of NLTE [Sr/Fe] vs. [Fe/H] also shrinks on the x-axis due to higher NLTE [Fe/H] ratios compared to LTE [Fe/H]. As presented in \figurename~\ref{metallicity_distribution_function}, it is evident that the metallicity distribution function below [Fe/H]$=-1.5$ drastically differs between LTE and NLTE: the NLTE [Fe/H] distribution shrinks, and we only probe the halo down to [Fe/H]$=-2.7$, due to increasing NLTE correction with decreasing [Fe/H].

In the middle-top panels of \figurename~\ref{Abund_Ba_Sr_Y}, we see that the LTE [Y/Fe] ratio shows a concave shape with [Fe/H], and on average sub-solar ($\langle\mathrm{[Y/Fe]}\rangle=-0.09$), which is consistent with past LTE studies (e.g. \citealt{delgado_mena_2017}). In contrast, NLTE [Y/Fe] ratios slightly decrease with increasing [Fe/H], from roughly $0.1$ at [Fe/H] $\lesssim -1.2$ to $\sim -0.1$ for [Fe/H] $\gtrsim -1.2$. We notice that the overall star-to-star [Y/Fe] scatter is equal to 0.17\,dex in both LTE and NLTE; this scatter slightly increases with [Fe/H] and ranges from 0.14 to 0.18 in LTE and 0.12 to 0.17 in NLTE. Such increase is likely due to the presence of [Y/Fe]-rich/poor stars (see below). On average, NLTE [Y/Fe] ratios are higher than LTE [Y/Fe] by $\sim 0.15\,$dex at [Fe/H] $\approx -2$, while the difference between NLTE and LTE decreases to 0.04 for [Fe/H] $\gtrsim -1$, which is similar but less pronounced than [Sr/Fe]. The overall behaviour of NLTE effects in Y with metallicity is consistent with the results of  \citet{Storm2023}.

In the top-right panel of \figurename~\ref{Abund_Ba_Sr_Y}, we show that LTE and NLTE [Ba/Fe] abundances follow a similar, although not the same, trend. As metallicities above $\sim -1$, NLTE [Ba/Fe] is slightly lower than LTE [Ba/Fe] by about 0.07\,dex, consistent with past studies, e.g. our previous work in \citet{Gallagher2020}, as we expect Ba lines equivalent widths to increase due to NLTE effects. NLTE [Ba/Fe] is rather flat for [Fe/H] $\gtrsim -1.25$, in agreement with previous studies (e.g. \citealt{delgado_mena_2017} in LTE and \citealt{korotin_2011} in NLTE).
The star-by-star standard deviation of NLTE [Ba/Fe] abundances is 0.16 dex, while it reaches 0.21 dex for LTE [Ba/Fe]. This shows that NLTE abundances have less intrinsic scatter, which has implications for the chemical enrichment of the elements in the Galaxy, as we will show in Sect.~\ref{sec:models} below.

\begin{figure}
\centering
\includegraphics[width=\columnwidth]{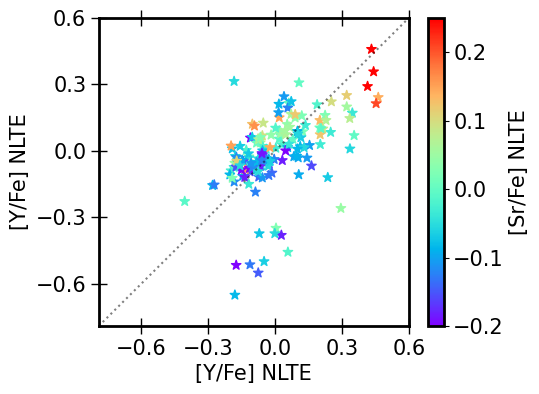}
\caption{NLTE chemical abundances of [Y/Fe] as a function of [Ba/Fe], color-coded with [Sr/Fe].}
\label{Ba_Y_Sr_Age}
\end{figure}

We notice the presence of stars deficient in [Y/Fe], compared to the main distribution at a given metallicity, as well as some stars with large [Ba/Fe] and [Y/Fe] values. To search for possible correlations between these low- and high-abundance stars, we present in \figurename~\ref{Ba_Y_Sr_Age} the NLTE abundances of [Y/Fe] as a function of [Ba/Fe], colour-coded with [Sr/Fe]. Firstly, we see that stars deficient in [Y/Fe] show also both sub-solar [Ba/Fe] and [Sr/Fe]. Secondly, stars with large [Y/Fe] ratios are also characterized by large [Ba/Fe] and [Sr/Fe]. These enhanced(deficient) stars do not seem to be preferentially part of any given Milky Way component,  when following the thin/thick disc and halo classification of \citet{Ruchti2011}. We also checked that for dwarfs and turn-off stars, there was no correlation with stellar ages from \citet{Serenelli2013}.

\begin{figure*}
\centering
\includegraphics[width=\textwidth]{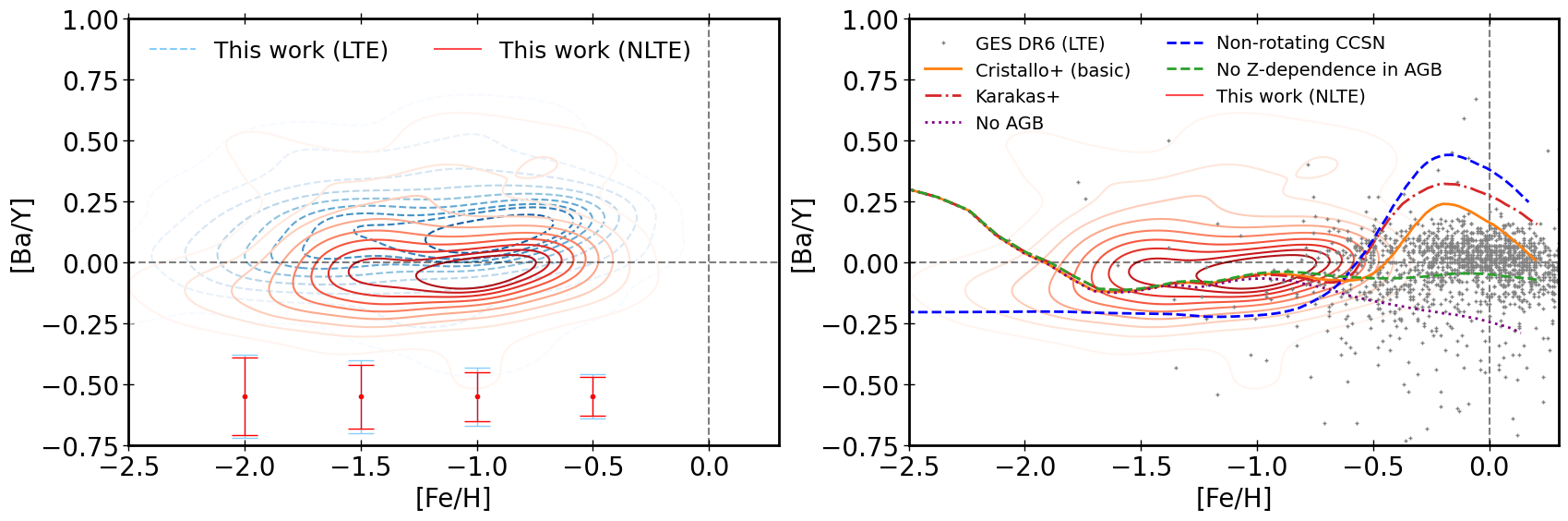}
\caption{Left panel: chemical abundance ratios [Ba/Y] in both LTE (blue contours) and NLTE (red contours) as a function of [Fe/H] (LTE and NLTE, respectively), for 187 stars. Error bars correspond to mean uncertainties in LTE and NLTE. Right panel, we only show NLTE [Ba/Y] (red contours) together with \emph{Gaia}-ESO sample used by \citet{Lian2023}, as well as 1-zone chemical evolution models with different yields prescriptions.}
\label{fig:Y_over_Ba}
\end{figure*} 
\section{Confronting chemical evolution model with heavy-to-light s-process element ratios}\label{sec:models}

We investigate here the [Ba/Y] ratio, which is a proxy of the heavy to light s-process elements \citep[e.g.][]{Lian2023}. We compare our observations to simple one-zone and inhomogeneous (stochastic) Galactic chemical evolution models, the former primarily relevant to understanding the chemical enrichment of the disc and latter qualitatively consistent with the present understanding of the formation of the Galactic halo (see for instance \citealt{Matteucci2021} and references there-in). 

In the left panel of \figurename~\ref{fig:Y_over_Ba}, we present LTE (blue contours) and NLTE (red contours) distributions of [Ba/Y] ratios for the 187 stars of our sample with available Ba and Y abundances (in both LTE and NLTE). Both LTE and NLTE [Ba/Y] abundances slightly increase with [Fe/H], and present both a similar scatter of 0.17\,dex, rather constant with [Fe/H]. Overall, We see that [Ba/Y] presents a weak dependence on [Fe/H] over the entire range of [Fe/H]. The main difference is a rigid shift of $-0.12$ dex between NLTE and LTE [Ba/Y]. The intrinsic scatter of abundance ratios is larger than the individual observational uncertainties of the measurement (ranging from 0.16 to 0.08 in NLTE, see Sect.~\ref{sec:data_methodology}) suggesting that the observational scatter is a signature of chemical enrichment processes. In the right panel of \figurename~\ref{fig:Y_over_Ba}, we show NLTE [Ba/Y] ratio as a function of NLTE [Fe/H]. We also overplotted the \emph{Gaia}-ESO NLTE [Ba/Y] abundances used in \citet{Lian2023}. Even though the abundances between this work and the Gaia-ESO were computed with different spectral analysis pipelines, the data is rather complementary\footnote{For the present study and the Gaia-ESO survey used the same linelist and model atmosphere grids.}.
\subsection{One-zone GCE models}

As in \citet{Lian2023}, we made of the OMEGA+ Galactic
chemical evolution (GCE) model \citep{Cote2017, Cote2018}, which includes gas inflow and outflows. The basic model includes core collapse supernova (CCSN) yields from \citep{limongi2018}, as well as Type Ia supernovae with yields from different Chandrasekhar- and sub-Chandrasekhar mass explosions as described in \citet{Eitner2023}. The GCE model also includes AGB yields from \citet{cristallo2015}. This basic GCE model is displayed in orange in \figurename~\ref{fig:Y_over_Ba}. For comparison, we also display a GCE model that includes recent AGB yields by \citet{Karakas2010}, which account for n-capture elements nucleosynthesis for metallicites down to $-2$ (see \citealt{Cinquegrana2022} and references there-in for more details). We also show one GCE model with no AGB contribution (in purple), and one with metallicity-independent AGB yields (in green). Finally, we added a GCE model that only includes non-rotating massive stars.

\citet{Lian2023} concluded that the shape of [Ba/Y] in the metal-rich regime ([Fe/H] $\gtrsim -0.6$) is driven by the metallicity dependence in the neutron capture efficiency in AGB stars. The mismatch between the GCE models and observations in this [Fe/H] regime is likely due to an overestimation of the s-process efficiency of low mass AGB stars \citep{Magrini2021}.

In the metallicity range -2 $\lesssim$ [Fe/H] $\lesssim -0.8$, the model without massive star yields shows a strong underproduction of [Ba/Y] and it is not consistent with the data. The large scatter in [Ba/Y] at a given metallicity could be a sign of chemical enrichment from AGBs with masses between 2 to 6 solar masses \citep[][their Fig. 5]{Lian2023}. We also find that the GCE track computed with non-rotating CCSN yields underpredicts our observations, which supports evidence from  the literature that CCSN resulting from the evolution of rapidly rotating massive stars are important sources of s-process elements (e.g \citealt{limongi2018}). In conclusion, such chemical evolution models with different nucleosynthesis prescriptions are not able to reproduce the data, mainly due to the large scatter in [Ba/Y] at a given [Fe/H].
\subsection{Stochastic GCE models}

Substantial efforts have been made in the GCE model community in trying to reproduce the abundance patterns measured in the Galactic halo stars. \citet{Cescutti2014} and \citet{Rizzuti2021} developed stochastic GCE models. Such models are meant to reproduce the chemical evolution of the Galactic halo, implying a series of nucleosynthesis events, overall on a time scale of 1 Gyr. Their model MRD+s B2 includes r-process contribution from magneto-rotational (MRD) supernovae and s-process nucleosynthesis from two channels: low-mass AGB stars and rotating massive stars (see \citealt{Cescutti2014} for more details). We adopted the model MRD+s B2. We explore here whether such models can reproduce the scatter we measure in our observations of [Ba/Y]. We notice that contrary to the model of \citet{Lian2023}, the models from \citet{Cescutti2014} do not include neutron star mergers.

\begin{figure}
\centering
\includegraphics[width=\columnwidth]{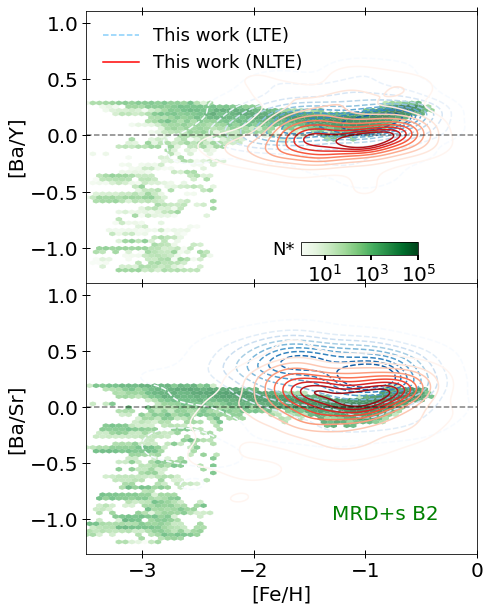}
\caption{[Ba/Y] (top) and [Ba/Sr] ratios as a function of [Fe/H]. Blue and red contours show LTE and NLTE abundances, respectively. We also show stochastic chemical evolution models (MRD+s B2) from \citet{Cescutti2014}.}
\label{fig:stochastic_BaY_BaSr_B2}
\end{figure}

In the top panel of \figurename~\ref{fig:stochastic_BaY_BaSr_B2}, we present LTE and NLTE [Ba/Y] abundance ratios as a function of NLTE [Fe/H] (blue and red contours, respectively). Additionally, we present stochastic GCE (colour-coded with number of stars in the model, i.e. SFR tracer). For [Fe/H]$\lesssim-2.5$, the GCE model shows a large scatter in the [Ba/Y] ratios, ranging from $\sim -1$ to $+0.3$. Such behaviour is directly attributed to the stochastic sampling of the IMF  during  the phase of halo formation \citep{Cescutti2014}. The predicted [Ba/Y] trend flattens out for metallicities [Fe/H]$\gtrsim-2.5$ and slightly rises up to [Ba/Y] $\approx 0.25$ for [Fe/H]$\gtrsim -1$. This GCE model matches rather well our LTE [Ba/Y] ratios, however, it overpredicts [Ba/Y] when compared to NLTE [Ba/Y]. This mis-match is simply the consequence of the fact that in this GCE model the yttrium yields were modified to match the LTE observations. Specifically the r-yields were scaled the LTE pattern of r-process rich stars. This pattern differs by a factor of 3 for Y, whereas it is consistent with the r-process solar residual for the remaining elements (see \citealt{Cescutti2014} for more details). Hence the GCE matches rather well our LTE [Ba/Y] pattern. Such results show that taking into account NLTE effects is key if one wants to accurately compare GCEs and observations. 

In the bottom panel of \figurename~\ref{fig:stochastic_BaY_BaSr_B2}, we show similar plots but for the [Ba/Sr] ratios. The GCE model shows a similar trend with [Ba/Y]. Contrary to [Ba/Y], the GCE model matches better the NLTE [Ba/Sr] ratios, while the LTE [Ba/Sr] ratios are clearly over-abundant compared to the GCE model. This could be due to the large NLTE effect when measuring the Sr line at 4607\AA. Again, taking properly into account NLTE effects is key when comparing chemical evolution models to observations, especially in the metal-poor regime.

\begin{figure}
\centering
\includegraphics[width=\columnwidth]{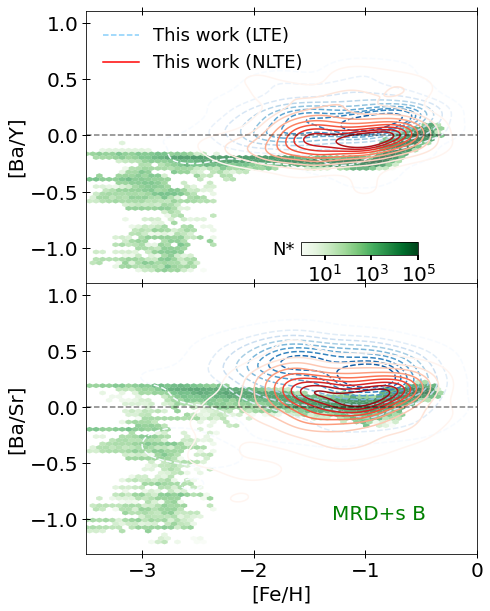}
\caption{Same figure as \figurename~\ref{fig:stochastic_BaY_BaSr_B2}, but we show the model MRD+s B from \citet{Cescutti2014}.}
\label{fig:stochastic_BaY_BaSr_B}
\end{figure}

Considering that we provide new NLTE abundances of neutron-capture elements for our stellar sample, it is necessary to confront our data to a stochastic chemical evolution model that is not scaled to reproduce the LTE abundance patterns. It corresponds to the model MRD+s B from \citet{Cescutti2014} which is similar to the model MRD+s B2, but the r-process contribution to Y production scaled to the solar residual, as the other elements and therefore not divided by a factor of 3. We show such a model in \figurename~\ref{fig:stochastic_BaY_BaSr_B}, together with the LTE and NLTE abundances of our metal-poor stars. Naturally, the predicted ratio of [Ba/Y] is lower by 0.3\,dex compared to the model MRD+s B2. Interestingly, the GCE model based on r-process solar residual is indeed closer to the NLTE observational distributions of [Ba/Y] and [Ba/Sr] in our stellar sample. It is important to remind the reader that our abundances are computed using 1D model atmospheres, and we expect the [Ba/Y] ratio to be lower when adopting an updated model atom (Storm et al. in prep) with updated Y+H collisional processes based on \citet{Wang2023}. Also, adopting 3D model atmospheres (Storm et al. in prep) can induce larger and more positive 3D NLTE effects for Y II lines. As a result, we do not see evidence for re-scaling the MRD  yields for Y to the pattern of the r-process rich stars as used in the model MRD+s B2 of \citet{Cescutti2014}; moreover, the r-process solar residuals appear to be more reliable, as expected: they are not affected by the NLTE corrections.

\section{Conclusions} \label{sec:conclusions}

In this letter, we focused on constraining the chemical
enrichment of [Ba/Fe], [Sr/Fe], [Y/Fe] and [Ba/Y] ratios of
Milky Way stars in the domain $-2.5\lesssim \feh \lesssim-0.5$
There is lack of abundance measurements in this [Fe/H] range
that deems direct constraints of Galactic chemical evolution
challenging.

\begin{enumerate}
    \item We used high-resolution (R$\in[35\,000-45\,000]$) and high signal-to-noise observations of RAVE metal-poor stars with previously determined $\teff$, $\logg$, and $\feh$ in both LTE and NLTE \citep{Ruchti2013}.
    
    \item We measured 1D LTE and NLTE abundances of [Ba/Fe], [Sr/Fe], and [Y/Fe] with associated uncertainties using the spectral synthesis code TSFitPy, a wrapper for NLTE version of TurboSpectrum. Careful visual checks of the spectral fits were performed in order to ensure the robustness of the determined chemical abundances. 
    
    \item We showed that the bulk of NLTE [Ba/Fe] ratios decreases with increasing [Fe/H], while NLTE [Sr/Fe] is rather constant with [Fe/H], and NLTE [Y/Fe] decreases with [Fe/H]. The combined NLTE effects are of the order of $-0.07$\,dex for [Ba/Fe], $+0.18$\,dex for [Sr/Fe], and $+0.10$\,dex for [Y/Fe] that also includes the NLTE effect on [Fe/H].
    
    \item Focusing on NLTE abundances, we find that stars enhanced(deficient) in [Ba/Fe] and [Y/Fe] are enhanced(deficient) in [Sr/Fe]. 
    
    \item We showed that the NLTE ratios of [Ba/Y] are centred around solar values, and the behaviour of the trend is rather flat with NLTE metallicity [Fe/H], implying that [Ba/Y] is not sensitive to [Fe/H]. The star-to-star scatter is substantial and is of the order of 0.2\,dex. LTE [Ba/Y] ratios show a similar dispersion as NLTE [Ba/Y], but are shitted by $+0.12$ dex relative to the solar values even at metallicities close to solar.
    
    \item  Single-zone chemical evolution models are unable to reproduce the [Ba/Y] scatter observed at a given metallicity. Such a scatter can, however, be more captured by stochastic GCE models. 
    
    \item Most importantly, we find a better agreement of the NLTE abundance ratios of light (Sr,Y) and heavy (Ba) element ratios compared to the GCE tracks from stochastic chemical evolution models with r-yields scaled r-process solar residual. 
    There is not anymore need of modifications of the r-process yields of yttrium following the pattern of r-process rich stars as was done e.g. in \citet{Cescutti2014}. Therefore, we conclude that properly taking into account NLTE effects when measuring abundances directly impacts comparisons between galactic chemical evolution models and observations, hence on yield prescriptions and nucleosynthesis channels.

\end{enumerate}

The current studied sample is still rather limited in the number of stars and in the metallicity coverage. Indeed, going to low-metallicity, typically down to [Fe/H] $\sim -4$ or  $-5$ will allow us to probe in more details the early neutron-capture enrichment of the Milky Way. In the near future, 4MOST will deliver hundreds of thousands of high-resolution (R$\sim$20000) optical spectra of the Milky Way disc, halo and bulge stars\citep{4MOST, bensby2019, Christlieb2019}, and will open a new era for NLTE abundances exploration down to very low metallicities. Strong effort will have to be put in order to provide the community with precise and accurate neutron-capture abundances from such a facility.

\begin{acknowledgements}
G.G. acknowledges support by Deutsche Forschungsgemeinschaft (DFG, German Research Foundation) – project-IDs: eBer-22-59652 (GU 2240/1-1 "Galactic Archaeology with Convolutional Neural-Networks:
Realising the potential of Gaia and 4MOST"). We acknowledge support by the Collaborative Research Centre SFB 881 (projects A5, A10), Heidelberg University, of the Deutsche Forschungsgemeinschaft (DFG, German Research Foundation) and by the European Research Council (ERC) under the European Union’s Horizon 2020 research and innovation programme (grant agreement 949173). M.B. is supported through the Lise Meitner grant from the Max Planck Society. This research was supported by the Munich Institute for Astro-, Particle and BioPhysics (MIAPbP) which is funded by the Deutsche Forschungsgemeinschaft (DFG, German Research Foundation) under Germany´s Excellence Strategy – EXC-2094 – 390783311. A.S. acknowledges grants PID2019-108709GB-I00 from Ministry of Science and Innovation (MICINN, Spain), Spanish program Unidad de Excelencia Mar\'{i}a de Maeztu CEX2020-001058-M, 2021-SGR-1526 (Generalitat de Catalunya). A.S. and G.C. acknowledge support from ChETEC-INFRA (EU project no. 101008324). GC acknowledges the grant PRIN project n.2022X4TM3H "Cosmic POT" from Ministero dell'Universit\'a e la Ricerca (MUR).
\end{acknowledgements}

\bibliographystyle{aa}        
\bibliography{paper}      

\appendix

\end{document}